\documentclass[fleqn,usenatbib,a4paper]{mnras}

\usepackage{comment}

\usepackage{newtxtext,newtxmath}
\usepackage[ruled,vlined]{algorithm2e} 
\usepackage[T1]{fontenc}

\DeclareRobustCommand{\VAN}[3]{#2}
\let\VANthebibliography\thebibliography
\def\thebibliography{\DeclareRobustCommand{\VAN}[3]{##3}\VANthebibliography}

\usepackage{graphicx}	%
\usepackage{subfig}
\usepackage[countmax]{subfloat}
\usepackage{amsmath}	
\usepackage{widetext}%
\usepackage{amssymb}
\usepackage{color,units}
\usepackage[dvipsnames]{xcolor}
\usepackage{float}
 \usepackage{capt-of}

\title[]{Inferring the population properties of binary black holes \\
from unresolved gravitational waves}

\author[Rory J.E. Smith et al.]{
Rory J. E. Smith,$^{1,2}$\thanks{E-mail: rory.smith@ligo.org (RJES)}
Colm Talbot,$^{3,1,2}$
Francisco Hernandez Vivanco$^{1,2}$
\and and Eric Thrane$^{1,2}$
\\
$^{1}$School of Physics and Astronomy, Monash University, Vic 3800, Australia\\
$^{2}$OzGrav: The ARC Centre of Excellence for Gravitational Wave Discovery, Clayton VIC 3800, Australia\\
$^{3}$LIGO, California Institute of Technology, Pasadena, CA 91125, USA
}

\pubyear{2020}

\begin{document}

\pagerange{\pageref{firstpage}--\pageref{lastpage}}
\maketitle

\begin{abstract}
The vast majority of compact binary mergers in the Universe produce gravitational waves that are too weak to yield unambiguous detections; they are unresolved.
We present a method to infer the population properties of compact binaries---such as their merger rates, mass spectrum, and spin distribution---using both resolved and unresolved gravitational waves.
By eliminating entirely the distinction between resolved and unresolved signals, we eliminate bias from selection effects.
To demonstrate this method, we carry out a Monte Carlo study using an astrophysically motivated population of binary black holes.
We show that some population properties of compact binaries are well constrained by unresolved signals after about one week of observation with Advanced LIGO at design sensitivity.
\end{abstract}

\maketitle


\section{Introduction}
Every year, around $2\times 10^6$ binary neutron stars and $1.5\times10^5$ binary black holes merge somewhere in the Universe, radiating gravitational waves \cite{GW170817_stoch}.
Only a small fraction of these signals are detected by observatories such as Advanced LIGO (aLIGO), Advanced Virgo, and KAGRA \cite{adv,aligo,kagra}.
The rest are too faint to be resolved.
Nonetheless, the ensemble of unresolved gravitational-wave signals forms an astrophysical background, which \textit{can} be detected by advanced gravitational-wave detectors~\cite{GW170817_stoch,GW150914_stoch,tbs,tbs-bns}.
Here, we use the word ``background'' to denote  gravitational-wave signals that are not clearly detected and published in catalogs, e.g.,~\cite{gwtc-1}.
Since there are many connotations associated with the notion of a gravitational-wave background, it is worth pausing to make our meaning absolutely clear.

First, we note that this definition of ``background'' is detector-dependent; as gravitational-wave detectors become more sensitive, a greater fraction of binary mergers will be clearly resolved, and so what we might refer to as background now will become foreground in the future.
Second, we note that the gravitational-wave background from compact binaries is often thought of as a {\em foreground} when looking for primordial gravitational waves from the early Universe; see, e.g.,~\cite{Maggiore}.
Indeed, one scientist's foreground is another's background; here we use the word ``foreground'' to refer to resolved binaries.
Finally, there is a common notion that the gravitational-wave background consists of a plethora of unimaginably faint sources.
In reality, it derives from a continuum of binaries, ranging from the nearly-detectable to the clearly-not-detectable.
Since there is no universally accepted definition of ``detection,'' the boundary between the resolved catalog and the unresolved background is fuzzy.

However one may choose to delineate this boundary, the background encodes rich information about the mass and spin distributions of compact binaries.
These distributions, in turn, provide insights into binary evolution~\cite{Stevenson,Stevenson2,salvo,spin,GerosaBerti,FarrNature,Wysocki18,eccentricity}, star formation history, the fate of massive stars~\cite{mass_uc,T&T18,O2R&P}, the behavior of matter at supranuclear densities \cite{PhysRevLett.121.161101}, and the existence of primordial black holes \cite{1475-7516-2017-09-037}, amongst other things.
Crucially, the foreground probes only the closest binaries.
By analyzing the foreground and background together, it is possible to probe the entire population of binary mergers.

Here, we use hierarchical inference\footnote{For a review of hierarchical inference in gravitational-wave astronomy, see Section~V of~\cite{ThraneTalbot}.} to extend the method outlined in~\cite{tbs} in order to  determine the ensemble properties of compact binaries.
By eliminating the artificial distinction between foreground and background, we probe greater distances than possible with resolved events alone, while eliminating bias from selection effects.
We demonstrate that it is possible to make population inferences even when excluding statistically significant, ``gold-plated'' detections.
The key results are posterior probability distributions describing the shape of the binary black hole mass and spin distributions, derived using entirely unresolved events.
We show that these posteriors are consistent with the true values used for the generation of the simulated data.
We argue that this method is statistically optimal in the sense that is not possible to obtain more narrow posteriors given a fixed dataset.

This work builds on~\cite{Gaebel}, which describes how population studies can be extended to include sub-threshold candidate events, some of which are bona fide gravitational-wave signals, even though any single candidate is probably a noise fluctuation.
This is part of a broader trend in gravitational-wave astronomy.
For example, the arguably marginal event GW170729 was included\footnote{The event GW170729 has an astrophysical probability ranging from $p_\text{astro}=48-98\%$.} in the first gravitational-wave transient catalog GWTC-1~\cite{O2} and the companion paper~\cite{O2R&P}.

We highlight a few innovations unique to this work.
First, we eliminate selection effects entirely by making no distinction between detected events and sub-threshold events.
Taking into account selection effects in population studies can be a somewhat subtle endeavour~\cite{ThraneTalbot,O1_bbh,Maya2,MandelFarrGair}, involving challenging efficiency calculations~\cite{Ng,Tiwari}.
These challenges are removed by eliminating the concept of a detection threshold.
Second, by eliminating the minimum detection threshold entirely, we extend the range of the analysis to include events at large redshifts, well beyond what can be probed with unambiguous detections.
This is an important first step toward studying the evolution of binary populations over cosmic time, though, more work is required to measure this redshift-dependence using hyper-parameters; see~\cite{Maya2}.
Third, while~\cite{Gaebel} generates pseudo posterior samples from a Fisher matrix approximation for the likelihood function, we calculate posterior samples using a full-fledged parameter estimation pipeline.
By carrying out full parameter estimation (the main computational cost of the search), we show that our method is computationally feasible.

The remainder of this paper is organized as follows.
In Section~\ref{sec:models} we describe astrophysically motivated models of the binary black hole mass spectrum and spin distributions.
In Section~\ref{sec:method}, we describe the method for population inference from a population of sub-threshold signals.
In Section~\ref{sec:results}, we present the results of our Monte Carlo study.
Concluding remarks are provided in Section~\ref{sec:conclusions}.

\section{Population Model}
\label{sec:models}
We parameterize the mass and spin distributions using one of the prescriptions from~\cite{O2R&P}.
In this section, we briefly summarize our population model.
The reader is referred to the appendix for more details.
Our models take the form of conditional priors $\pi_{\theta}(\theta|\Lambda)$ where $\theta$ are binary black hole parameters and $\Lambda$ are hyper-parameters governing the shape of the $\theta$ distribution.
A list of hyper-parameters, their meaning, and injection values used in this study is provided in Tab.~\ref{table:hyperparams}.

\begin{table*}
    \centering
    \begin{tabular}{c  c c}
        \hline\hline
        Hyper parameter $\Lambda_i$ &  Description & Injection value\\
        \hline
        $\xi$ & Astrophysical duty cycle & $6.67\times10^{-3}$\\
        $m_{\text{min}} (M_{\odot})$ & Minimum black hole mass & $8.68M_{\odot}$\\
        $m_{\text{max}} (M_{\odot})$ & Maximum mass of black holes in the power law component & $39.5 M_{\odot}$
\\
        $\mu_m (M_{\odot})$ & Mean of the Gaussian component of the primary mass distribution & $33.4 M_{\odot}$ 
 \\
        $\sigma_m$ & Standard deviation of the Gaussian component of the primary mass distribution & $1.08 M_{\odot}$
 \\
        $\lambda_m$ & Fraction of black holes in the Gaussian component of the primary mass distribution & $0.340$
\\
        $\alpha_m$ & Slope of the power law component of the primary mass distribution & 2.00
 \\
        $\beta_m$ & Slope of the mass ratio distribution & -0.198
 \\ 
        $a_{\text{max}}$ & Maximum spin magnitude & 1.00 
 \\
        $\alpha_{a}$ & Spin-magnitude beta distribution slope parameter (rise) & 1.50
 \\
        $\beta_{a}$ & Spin-magnitude beta distribution slope parameter (fall) & 3.50
 \\
        $\sigma_{\text{tilt}}$ & Standard deviation of the spin-tilt angle distribution & 1.00
\\
        $\xi_{\text{tilt}}$ & Fraction of BBHs with Guassian distributed spin tilts & 0.50
\\
        \hline\hline
    \end{tabular}
    \caption{Hyper parameters $\Lambda_i$ of the binary black hole mass and spin population distributions.}
    \label{table:hyperparams}
\end{table*}

\begin{figure}
\centering
\includegraphics[width=0.48\textwidth]{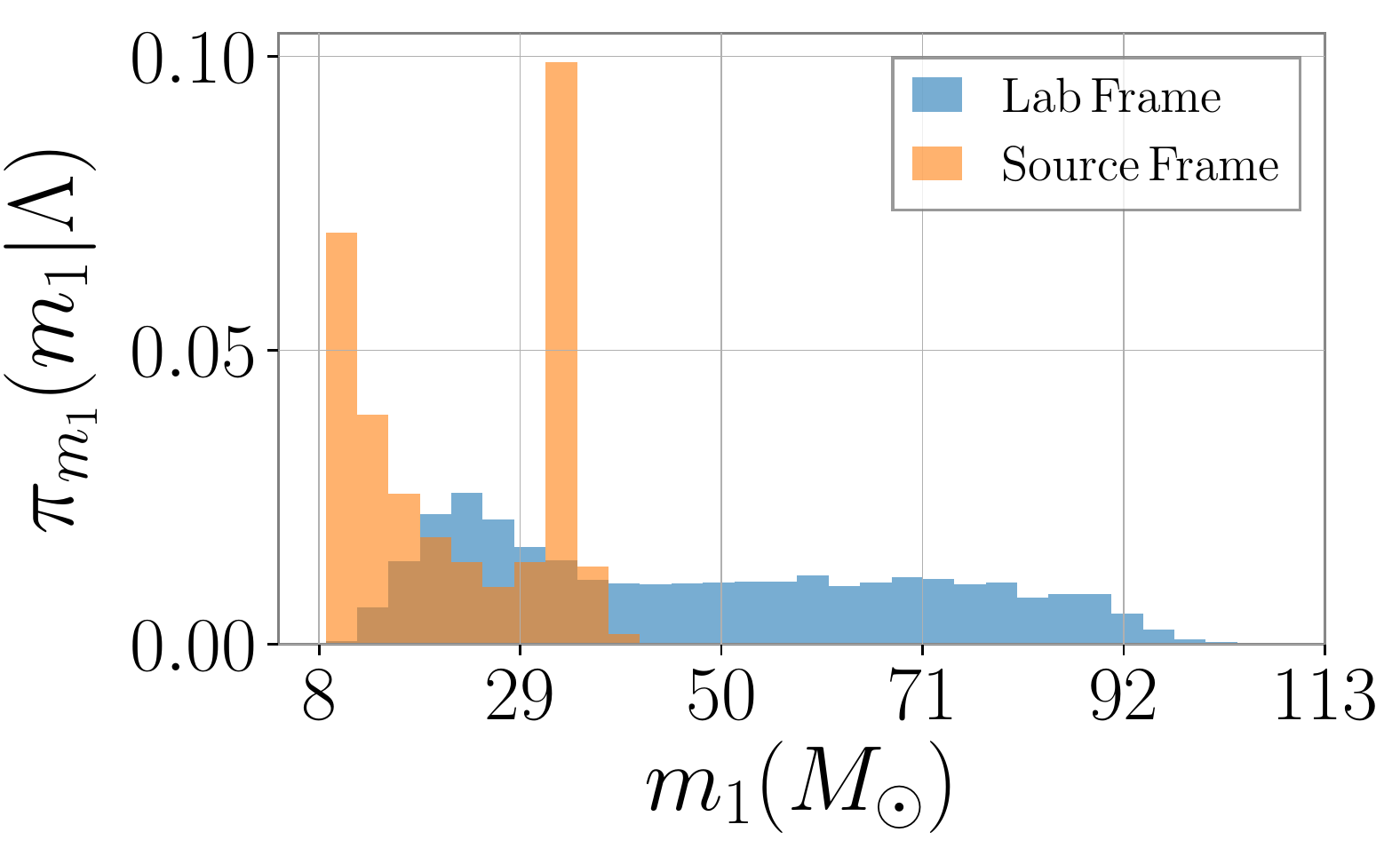}
\caption{Astrophysically motivated primary mass ($m_1$) distribution in the source frame (orange) and lab frame (blue). The lab-frame mass distribution appears redshifted due to the expansion of the universe.
}
\label{fig:redshifted_pop_mass}
\end{figure}

We model the black hole mass spectrum following~\cite{T&T18}.
The distribution is a mixture model of a truncated power-law and a Gaussian.
An example of the source-frame primary mass distribution is shown in orange in  Fig.~\ref{fig:redshifted_pop_mass} and the lab-frame distribution (distorted bt cosmological redshift) is shown in blue.
we model the distribution of black hole spin magnitudes following~\cite{Wysocki18}.
The distribution is a beta distribution.
We model the distribution of black hole spin orientations following~\cite{T&T17}.
The distribution is a mixture model of an isotropic distribution and model with a preference for aligned spin. For this study, we choose a set of plausible population parameters based on \cite{O2R&P}.

\begin{figure*}
\centering
\subfloat[]{
\label{fig:p_of_z}
\includegraphics[width=0.48\textwidth]{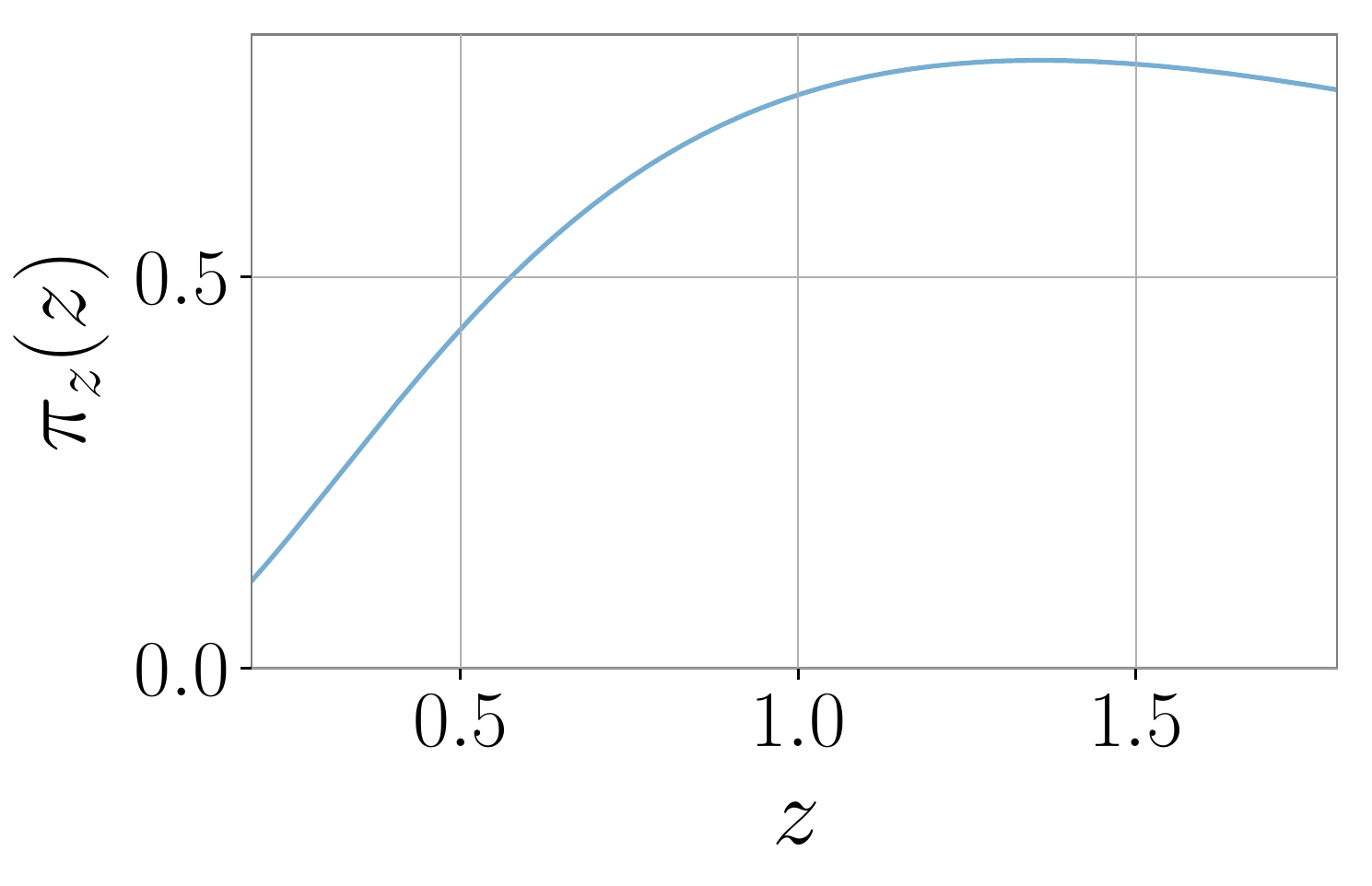}}
\subfloat[]{
\label{fig:p_of_dL}
\includegraphics[width=0.48\textwidth]{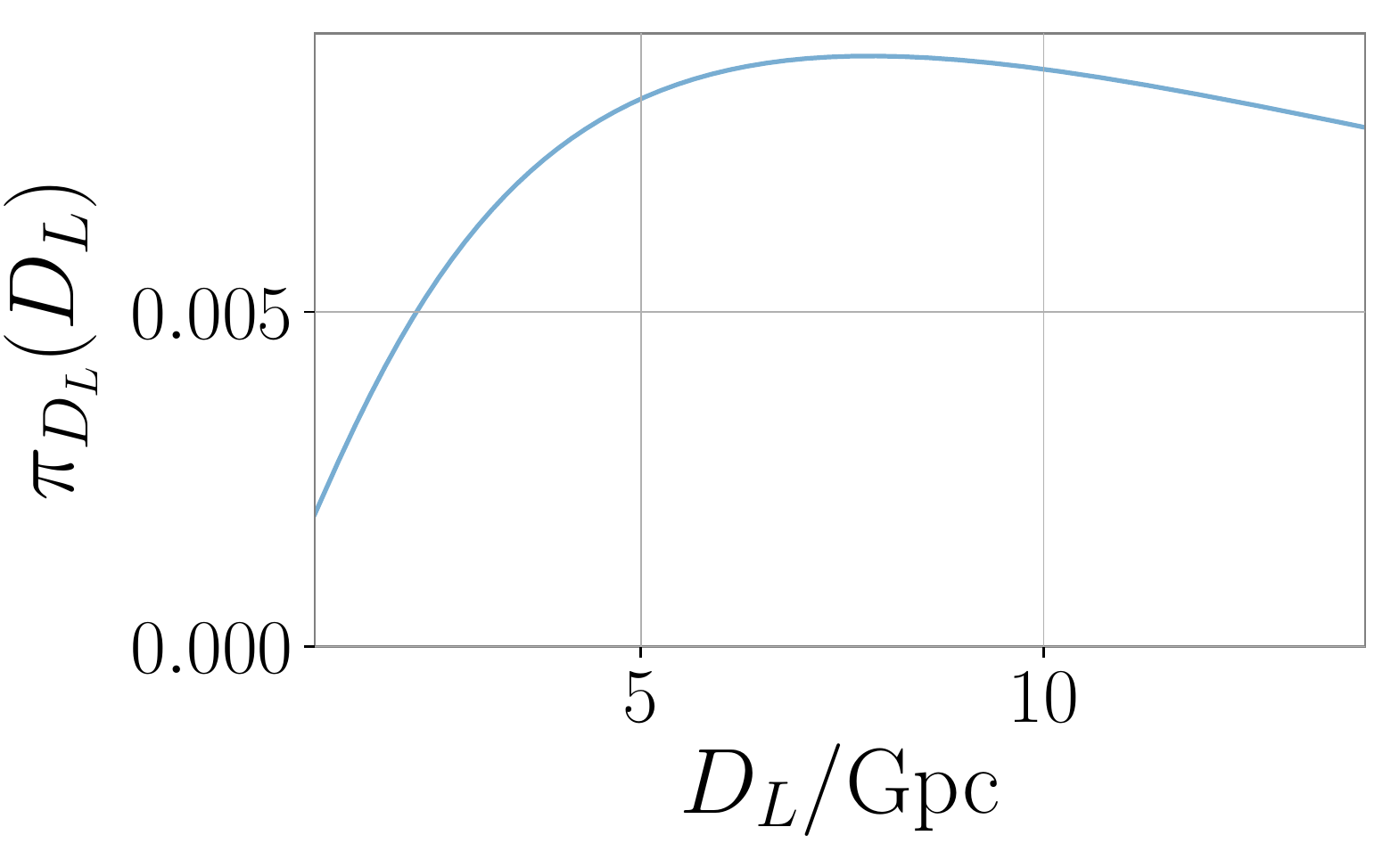}}
\centering
\caption{
Prior distributions on redshift (left) and luminosity distance (right)
}
\end{figure*}

We assume a fixed, known redshift distribution of (or equivalently, luminosity distance).
We assume that sources are uniformly distributed in co-moving volume to a maximum luminosity distance of $d_L^\text{max}\approx\unit[5]{Gpc}$ (redshift $z=0.8$).
Throughout, we assume the standard $\Lambda$CDM cosmology ($\Omega_\Lambda=0.69, \Omega_m=0.31, \unit[H_0=67.7]{km \,Mpc^{-1}s^{-1}}$) \cite{Planck_params}. 
While this distance distribution ignores effects arising from the time-dependent star-formation rate, see~\cite{Maya2,standard-siren}, it is satisfactory for our present purposes.
By probing redshifts up to $z=0.8$ (lookback time = $\unit[7]{Gyr}$), it is in-principle possible to glean information about a time when the Universe was  younger and the star formation rate was higher~\cite{madauSFR}.
In Fig.~\ref{fig:p_of_z} and Fig.~\ref{fig:p_of_dL} we show the explicit redshift and luminosity distributions implied by our uniform-in-comoving volume source distribution with standard $\Lambda$CDM cosmology.

The final ingredient required to characterize our population of binary black holes is the duty cycle $\xi$, the fraction of segments containing a binary black hole signal.
In the next section, we describe how the data are divided into $\unit[16]{s}$ segments.
Current observations of binary black hole mergers suggest that two black holes merge somewhere in the Universe on average once every $\unit[223^{+352}_{-115}]{s}$.
Most of these mergers probably take place at redshifts of $z<2$ ($d_L \lesssim \unit[15]{Gpc})$.
Beyond $z=2$, it is believed that star-formation rate decreases~\cite{madauSFR}.
With fewer stars, there are fewer black holes, and therefore fewer mergers.
Assuming an average time between binary black hole of $\unit[100]{s}$ out to $d_L = 15$ Gpc, the duty cycle out to luminosity distances of $\unit[5]{Gpc}$ is approximately $\xi=6.67\times10^{-3}$, and so we use this value for our injection study.

\section{Inferences from the gravitational-wave background}\label{sec:method}
\subsection{Overview}
This section describes the statistical formalism that allows us to calculate the hyper-posterior distribution $p(\Lambda|\vec{d})$ for population parameters $\Lambda$ described in Section~\ref{sec:models} given some dataset $\vec{d}$.
We follow the method described in~\cite{tbs}.
The calculation is divided into the following steps.
\begin{enumerate}
    \item We divide the data into $\unit[16]{s}$ segments.
    These segments are a convenient size so that any given segment is unlikely to contain more than one binary black hole signal. However, they are long enough that it is relatively unlikely for a binary black hole signal to fall on the boundary of two segments; see~\cite{tbs}.
    \item Run the nested sampling code \texttt{dynesty} \cite{dynesty} (implemented in the {\tt bilby}~\cite{bilby} Bayesian inference library) to generate posterior samples $\{\theta_{k,i}\}$ describing the mass and spins of individual binary black hole events in each segment. Additionally, {\tt dynesty} estimates for each data segment, the noise evidence ${\cal Z}(d|{\cal H}_N)$---that there is no binary black hole present---and ``the default signal evidence'' ${\cal Z}_{\O}(d|{\cal H}_S)$---that there is a binary black hole signal present given some default prior $\pi_{\O}(\theta)$.
    \item The posterior samples and evidences for each segments are used to define a ``total likelihood'', defined in Eq.~\ref{eq:combined_hyper_L}, which combines data from many segments.
    We discuss the hyper likelihood in greater detail in the next subsection.
    \item Having defined the hyper likelihood, we use \texttt{dynesty} to generate hyper-posterior samples $\{\Lambda_l\}$, which provide a representation of $p(\Lambda|\vec{d})$.
    
\end{enumerate}

Steps 1-2 are relatively straightforward.
In the next subsection, we describe the hyper likelihood used in steps 3-4.

\subsection{The hyper likelihood}
Following~\cite{tbs}, we employ a likelihood function to describe the probability of some large dataset $\vec{d}$ given a population of binary black hole described by hyper-parameters $\xi$ (the fraction of data segments containing a signal) and $\Lambda$, which describes the shape of the binary black hole mass and spin distributions
\begin{align}
\label{eq:combined_hyper_L}
\mathfrak{L}^\text{tot}\big(\vec{d}|\Lambda,\xi\big) = & \prod_i^n \bigg[ \xi \, {\cal L}(d_i|\Lambda,\mathcal{H}_S) + 
(1-\xi) {\cal Z}(d_i|\mathcal{H}_N) \bigg] .
\end{align}
There is a lot to explain in this equation and the rest of this subsection is devoted to this task.
The $\text{tot}$ superscript denotes that this is the likelihood for the entire dataset $\vec{d}$.
The expression includes a product over $i$ data segments running from $i=1$ to $n$.
The term ${\cal L}(d_i|\Lambda,\mathcal{H}_S)$ is the single-segment Bayesian evidence for the data $d_i$ in segment $i$ given the signal hypothesis $\mathcal{H}_S$ and hyper-parameters $\Lambda$.
The term ${\cal Z}(d_i|\mathcal{H}_N)$ is the single-segment noise evidence for the data $d_i$ in segment $i$.
The hyper-parameter $\xi$ is often referred to as ``duty cycle,'' and may be converted into a rate~\cite{tbs}.

The single-segment noise evidence ${\cal Z}(d_i|\mathcal{H}_N)$ is straightforwardly calculated for each segment using a Gaussian-noise likelihood\footnote{We note that this is missing a normalisation factor, however, as this only depends on the PSD and not on the template, we can freely factor this out of the both the signal and noise evidences.}
\begin{align}
    {\cal Z}(d_i|\mathcal{H}_N) =
    \exp\bigg(-\frac{1}{2}\langle d_i, d_i \rangle \bigg) .
\end{align}
Here, we employ a noise-weighted inner product
\begin{align}
    \langle a, b \rangle \equiv 4\Re\Delta f
    \sum_k \frac{a^{*}(f_k) b(f_k)}{S_n(f_k)} ,
\end{align}
where the sum is over frequency bins $k$ with bin widths of $\Delta f$ and $S_n(f)$ is the strain noise power spectral density.

The single-segment signal likelihood ${\cal L}(d_i|\Lambda,\mathcal{H}_S)$ is given by
(Eq.~\ref{eq:Z0}) yielding:
\begin{align}\label{eq:hyper}
{\cal L}(d_i|\Lambda,\mathcal{H}_S) \approx
\frac{\mathcal{Z}_{\O}(d_i|\mathcal{H}_S)}{n_s} \sum_{k=1}^{n_s}
\frac{\pi(\theta_{k,i}|\Lambda)}{\pi_{\O}(\theta_{k,i})} .
\end{align}
Here, $\mathcal{Z}_{\O}(d|\mathcal{H}_S)$ is the Bayesian evidence for a binary black hole signal in segment $i$ calculated using some default prior for the binary black hole parameters $\theta$ denoted $\pi_{\O}(\theta)$.
Assuming Gaussian noise, it is given by
\begin{align}
    \mathcal{Z}_{\O}(d_i|\mathcal{H}_S) \equiv & \int d\theta_i \, 
    {\cal L}(d_i|\theta,\mathcal{H}_S) 
    \pi_{\O}(\theta_i) \label{eq:Z0} \nonumber\\
    = & \int d\theta_i \,
    \exp\Big(-\frac{1}{2}\big\langle d_i-h(\theta_i), d_i-h(\theta_i)\big\rangle \Big) \nonumber\\ 
    &  \pi_{\O}(\theta_i),
\end{align}
where $h(\theta)$ is the gravitational waveform, in this case, calculated  {\tt IMRPhenomPv2} approximant ~\cite{IMRPhenomP,PhysRevD.94.044031}.
The integral in Eq.~\ref{eq:Z0} is calculated numerically using the Bayesian inference library, {\tt bilby}~\cite{bilby} implementation of {\tt dynesty}~\cite{dynesty}.
In addition to calculating $\mathcal{Z}_{\O}(d_i|\mathcal{H}_S)$, {\tt bilby} outputs a list of $n_s$ posterior samples $\{\theta_{k,i}\}$, which describe the posterior $p(\theta_i|d_i)$ given the default prior.
It is sometimes said that the ratio of priors $\pi(\theta_{k,i}|\Lambda)/\pi_{\O}(\theta_{k,i})$ in Eq.~\ref{eq:hyper} serves to ``reweight'' the posterior samples calculated using the default prior $\pi_{\O}(\theta)$~\cite{ThraneTalbot}.

\subsection{The hyper-posterior}
Using the hyper likelihood defined in Eq.~\ref{eq:combined_hyper_L}, it is straightforward to obtain the (hyper-) posterior for duty cycle and the other hyper-parameters $\Lambda$
\label{eq:hyper_posterior}
\begin{align}
p(\Lambda,\xi|\vec{d}) = \frac{\mathfrak{L}^\text{tot}\big(\vec{d}|\Lambda,\xi\big) \pi(\Lambda) \pi(\xi)}{\mathcal{Z}_{\Lambda}^{\text{pop}}} .
\end{align}
Here $\pi(\Lambda)$ is the hyper-parameter prior, which we take to be uniform for each hyper-parameter.
The distribution $\pi(\xi)$ is the duty cycle prior.
In a real analysis, one should choose a distribution, which uses a Poisson distribution to relate duty cycle to astrophysical rate; see~\cite{tbs}.
However, for our present purposes, it is convenient to simply employ a uniform prior.
The variable $\mathcal{Z}_{\vec{pop}}^{\Lambda}$ is the hyper-evidence.
They hyper-evidence can be used to carry out model selection between different population models; see~\cite{T&T17,T&T18,Stevenson,Stevenson2,O2R&P,Stevenson2,salvo,spin,GerosaBerti,FarrNature,Wysocki18,eccentricity}.

\section{Results: Demonstration with simulated data}
\label{sec:results}
We analyze $\unit[5.5]{days}$ of simulated Advanced LIGO (aLIGO) design-sensitivity data~\cite{aligo} containing an ensemble of 200 simulated binary black hole signals.
We divide the data into $3\times10^4$ sixteen-second segments.
This yields a duty cycle $\xi=200/30000 = 6.67\times10^{-3}$.
We derive the duty cycle by first assuming an average merger range of binary black holes of 1 per 100s. 
We then assume that the merger rate drops significantly beyond a redshift of $z\sim 2$ so that their contribution can be effectively ignored.
The fraction of all binaries contained in the volume with maximum redshift considered here, $z=0.8$, is approximately 4\%.
The average merger rate out to $z=0.8$ is then approximately one merger per 45min. In 5.5 days this yields 176 binary mergers, however we choose to round up to 200.

The masses and spins of the binary black hole's are drawn from the mass and spin distributions described in Sec.~\ref{sec:models}.
The remaining ``extrinsic'' parameters are drawn using standard distributions.
All of the signals in our injection set are below the usual threshold for matched-filter network SNR: $\rho_\text{network}^\text{th}=12$.
Based on results from~\cite{tbs}, we expect the binary black hole background to be detectable with approximately one day of aLIGO design sensitivity data.

We estimate the signal and noise evidence ${\cal Z}_S, {\cal Z}_N$, and obtain posterior samples for binary black hole source parameters for every data segment.
The priors, summarized in Table~\ref{table:runpriors}, and are chosen to be relatively uninformative so we can recycle the posterior samples later.
We then use the sets of evidence and posterior samples as input to Eq.~\ref{eq:hyper_posterior} to compute the posterior for $\Lambda$---the population mass and spin distribution parameters---and $\xi$, the astrophysical duty cycle.

\begin{table}
    \centering
    \begin{tabular}{c  c}
        \hline\hline
        Parameter $\theta_i$ & Prior $\pi(\theta_i)$ \\
        \hline
        $m_1$ & Uniform(6$M_{\odot}$,50$M_{\odot}$) \\
        $q$ & Uniform(0.2,1) \\
        $D_C^3$ & Uniform($1\text{Gpc}^3$, $5^3\text{Gpc}^3$)\\
        $t_c$ & Uniform($0$s, $16$s) \\
        $\cos\iota$ & Uniform($-1$,1) \\
        $\phi_c$ & Uniform(0,2$\pi$) \\
        $\psi$ & Uniform(0,$\pi$) \\
        $\cos t_1$ & Uniform(-1,1) \\         $\cos t_2$ & Uniform(-1,1) \\ 
        $\phi_{12}$ & Uniform(0,$2\pi$) \\
        $\phi_{JL}$ & Uniform(0,$2\pi$) \\
        $a_1$ & Uniform(0,1)\\
        $a_2$ & Uniform(0,1)\\
        $\alpha$ & Uniform(0,$2\pi$)\\
        $\cos\delta$ & Uniform(-1,1)\\
        \hline\hline
    \end{tabular}
    \caption{
    Priors on the 15 binary black hole signal parameters, $\pi(\theta)$.
    The priors are used in Stage 1 of the hierarchical population inference (Sec.~\ref{sec:method}).
    The parameters are the source-frame primary black-hole mass, $m_1$; mass ratio $q$; co-moving distance $D_C$; time of coalescence $t_c$; cosine of the orbital inclination $\cos\iota$; phase at coalescence $\phi_c$; polarization phase $\psi$; cosine of the spin-tilt angles $\cos t_1$ and $\cos t_2$; the angle between the two spin vectors $\phi_{12}$; angle between the total and orbital angular momentum $\phi_{JL}$; dimensionless spin magnitudes $a_1$ and $a_2$; right ascension $\alpha$; and cosine of the declination $\delta$.}
    \label{table:runpriors}
\end{table}

The computational cost of running full parameter estimation on $3\times10^4$ 16-second data segments is kept manageable by explicitly marginalizing over three parameters, which are difficult to sample: comoving distance, coalescence time, and coalescence phase; see e.g.,~\cite{ThraneTalbot} for the details of these marginalization schemes.
By marginalizing over these parameters, we significantly decrease the convergence time, and hence run time, of computing evidences and drawing posterior samples in step~1.

We find that the background is detectable within one week out to comoving distances of $\unit[5]{Gpc}$, assuming masses and spins drawn from the distribution described in Sec.~\ref{sec:models}.
The posterior distribution on $\xi$ is consistent with the true value of $\xi=0.67\%$, and the log Bayes factor (Eq.~15 of \cite{tbs}) overwhelmingly supports a detection of a population of compact binaries: $\ln \text{BF} \approx 700$, confirming the previous result from~\cite{tbs} with a different, more realistic population of BBH.

\begin{figure*}
\centering
\subfloat[]
{\label{fig:pop_mass}
\includegraphics[width=0.96\textwidth]{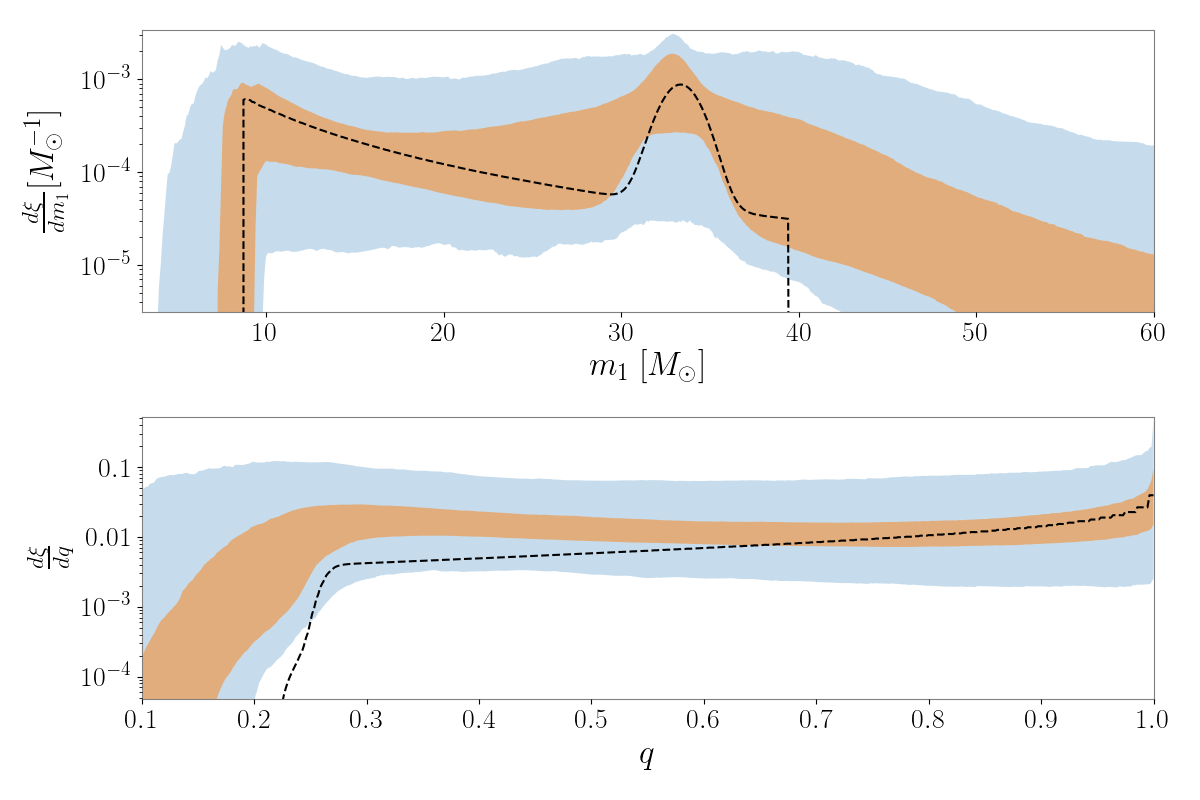}}
\quad
\subfloat[]{
\label{fig:pop_mags}
\includegraphics[width=0.48\textwidth]{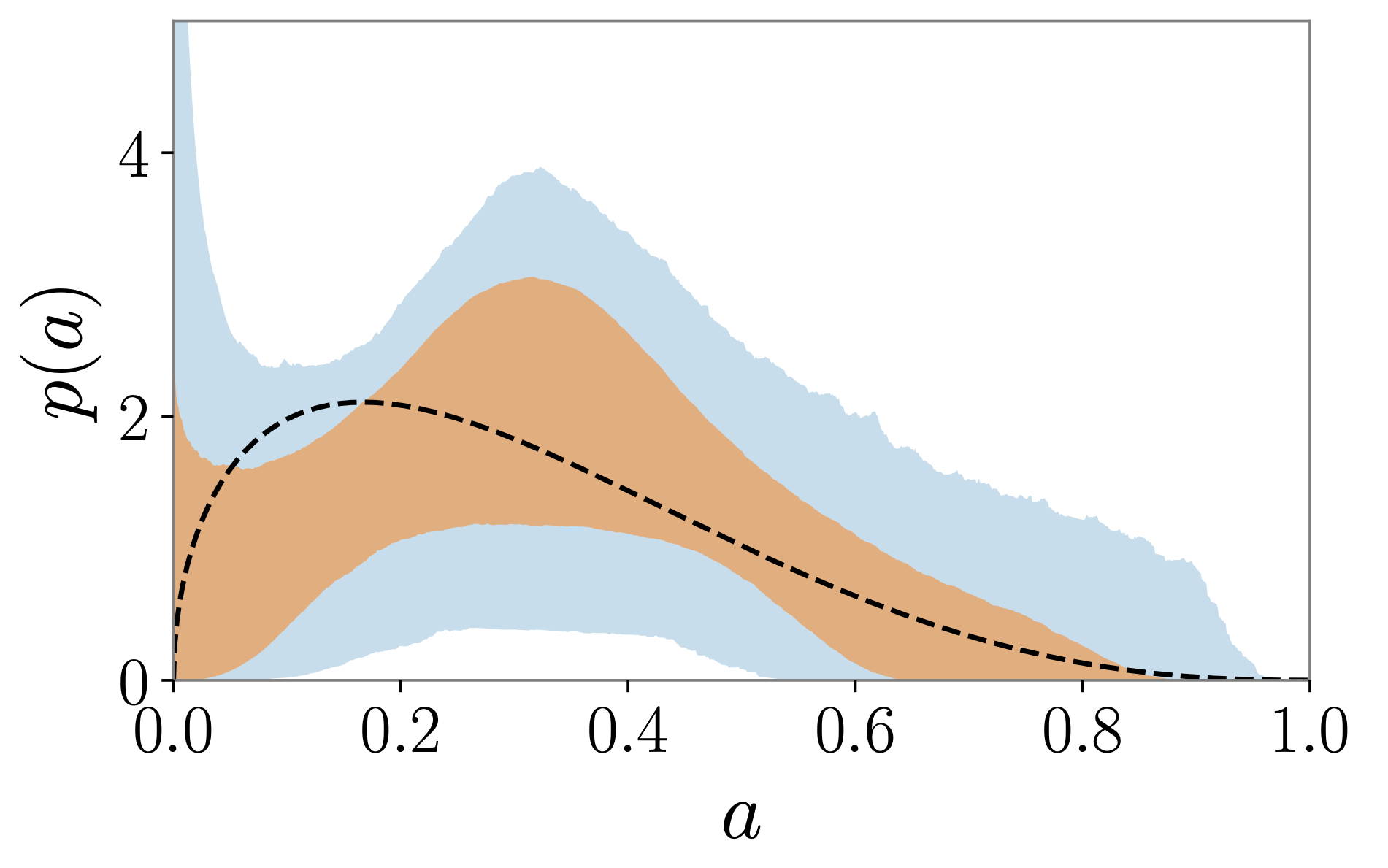}}
\subfloat[]{
\label{fig:pop_tilts}
\includegraphics[width=0.48\textwidth]{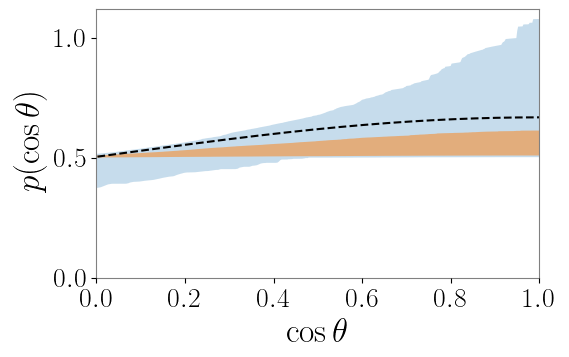}}
\centering
\caption{\label{fig:multi_panel} 
Posterior predictive distributions of binary black hole parameters.
These results are obtained using five and a half days of simulated aLIGO data containing 395 binary black holes signals.
The dashed line is the true distribution, while the red contours represent the 50$\%$ (light) and 90$\%$ credible intervals on the inferred distributions.
The parameters are: (top) primary black hole mass $m_1$, (center) mass ratio $q$, (lower left) spin magnitude $a$, (lower right) cosine spin tilt $\cos\theta$.
}
\end{figure*}

We find that we can begin to constrain some of the mass and spin population parameters are using the the 200 unresolved mergers in our simulated data.
In Fig.~\ref{fig:pop_mass} we show posterior predictive distributions for different mass and spin parameters.
The posterior predictive distributions reflect our updated prior based on information from our hyper-posteriors; see~\cite{ThraneTalbot}. 
The contours represent the $1-\sigma$ and $2-\sigma$ credible intervals.

In Fig.~\ref{fig:ppsn}, we show posterior distributions for hyper-parameters associated with the duty cycle and mass parameters.
In~\ref{fig:other}, we show posterior distributions for the parameters associated with the Gaussian component of the mass population model.
In Fig.~\ref{fig:spins}, we show posterior distributions for hyper-parameters describing black hole spins.

\section{How sensitive are we to subthreshold events?}
\begin{figure}[!t]
\centering
\includegraphics[width=0.48\textwidth]{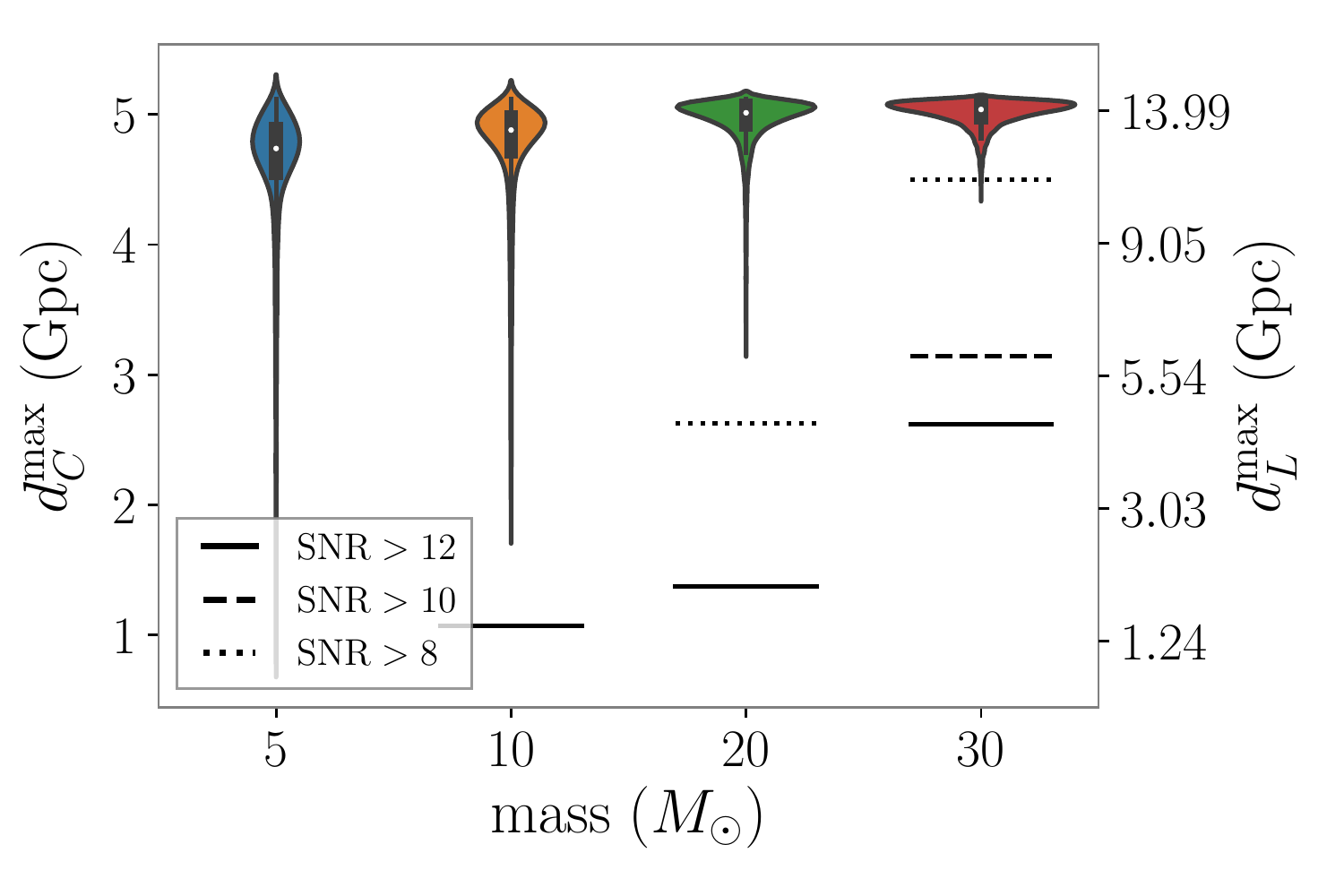}
\caption{Violin plots of the comoving and luminosity $d_\text{max}$ posterior obtained by running hierarchical inference with different mass distributions. Each distribution is set to be Gaussian with standard deviation $\sigma=0.3M_\odot$ and means $\mu = \left( 5, 10, 20, 30 \right) M_\odot$. The horizontal solid, dashed and dotted lines correspond to the most distant event observed with network SNR$>$12, SNR$>$10 and SNR$>$8 for each mass respectively. The posteriors peak at the true value $d_\text{max}=\unit[5100]{Mpc}$ (comoving distance) and the most distant events with SNR$>12$ lie below the $d_\text{max}$ posteriors, suggesting that we obtain most of the information from subthreshold events.
}
\label{fig:violin_plots}
\end{figure}

In this section, we investigate where the information for our analysis comes from.
Is our resolving power coming primarily from binaries just below the detection threshold, or do we gain information from weaker events as well?
To address this question, we carry out a follow-up study where we introduce a new hyper-parameter, $d_\text{max}$, the maximum comoving distance for binary mergers.  In our new population model, the rate of binary mergers drops to zero for distances greater than $d_\text{max}$. 
The $d_\text{max}$ parameter is not physical, but it is useful for our present investigation: if the data disfavor some value of $d_\text{max}$ (less than the true value of $d_\text{max}$), then we are getting information from that distance.
We set the true value of $d_\text{max}$ = $\unit[5100]{Gpc}$ (comoving distance) and then use hierarchical inference to obtain a posterior for $d_\text{max}$. We calculate the posterior on $d_\text{max}$ for different Gaussian mass distributions with standard deviation $\sigma=\unit[0.3]{M_\odot}$ and means $\mu = \left( 5M_\odot,10 M_\odot, 20M_\odot, 30M_\odot \right)$. The results are shown in Fig.~\ref{fig:violin_plots}.

The posterior on $d_\text{max}$ peaks at the true value of $d_\text{max}=\unit[5100]{Mpc}$ (comoving distance). 
The $d_\text{max}$ likelihood is clearly informative for distances greater than the distance of the furthest SNR$>$12 event, which is marked by the horizontal solid black line in Fig.~\ref{fig:violin_plots}. 
This is true for all masses considered in our study.
A similar conclusion is made for the most distant event with SNR$>$10, marked by the horizontal dashed line and the most distant event with SNR$>$8, marked the horizontal dotted line.
(No events with SNR$>$12 were used to obtain this hyper-posterior.) 
This plot is a good indication that we are indeed getting information from sub-threshold events.

\section{Conclusions}
\label{sec:conclusions}
Our results demonstrate that the astrophysical gravitational-wave background can be used to constrain the population properties of binary black holes together with ``gold plated'' foreground signals.
By applying hierarchical inference to all available data---irrespective of whether it contains a gravitational-wave signal or not---we eliminate selection bias. 
By carrying out population inferences with sub-threshold events we help extend the reach of the current generation of observatories to greater distances.
A crucial next step is the demonstration of the algorithm using real data.
A mock data challenge is underway to show how the algorithm performs in realistic conditions.
Another goal is to determine how much information can be inferred about the redshift dependence of binary-black hole mass and spin properties.

\section{Acknowledgements}
RS, CT, FHV, and ET are supported by the Australian Research Council (ARC) CE170100004.
ET is supported by ARC FT150100281.
We thank Stuart Anderson and the LIGO Data Grid for assistance with computing infrastructure, and Maya Fishbach, Thomas Callister and Thomas Dent for helpful comments and suggestions.
We acknowledge the OzStar cluster for providing graphical processor units to carry out some of our calculations.

\appendix
\section{Population hyper parameter estimation}
The one and two dimensional PDFs for the population hyper parameters used in this study are shown below. 
\begin{figure*}[b]
    \includegraphics[width=1\textwidth]{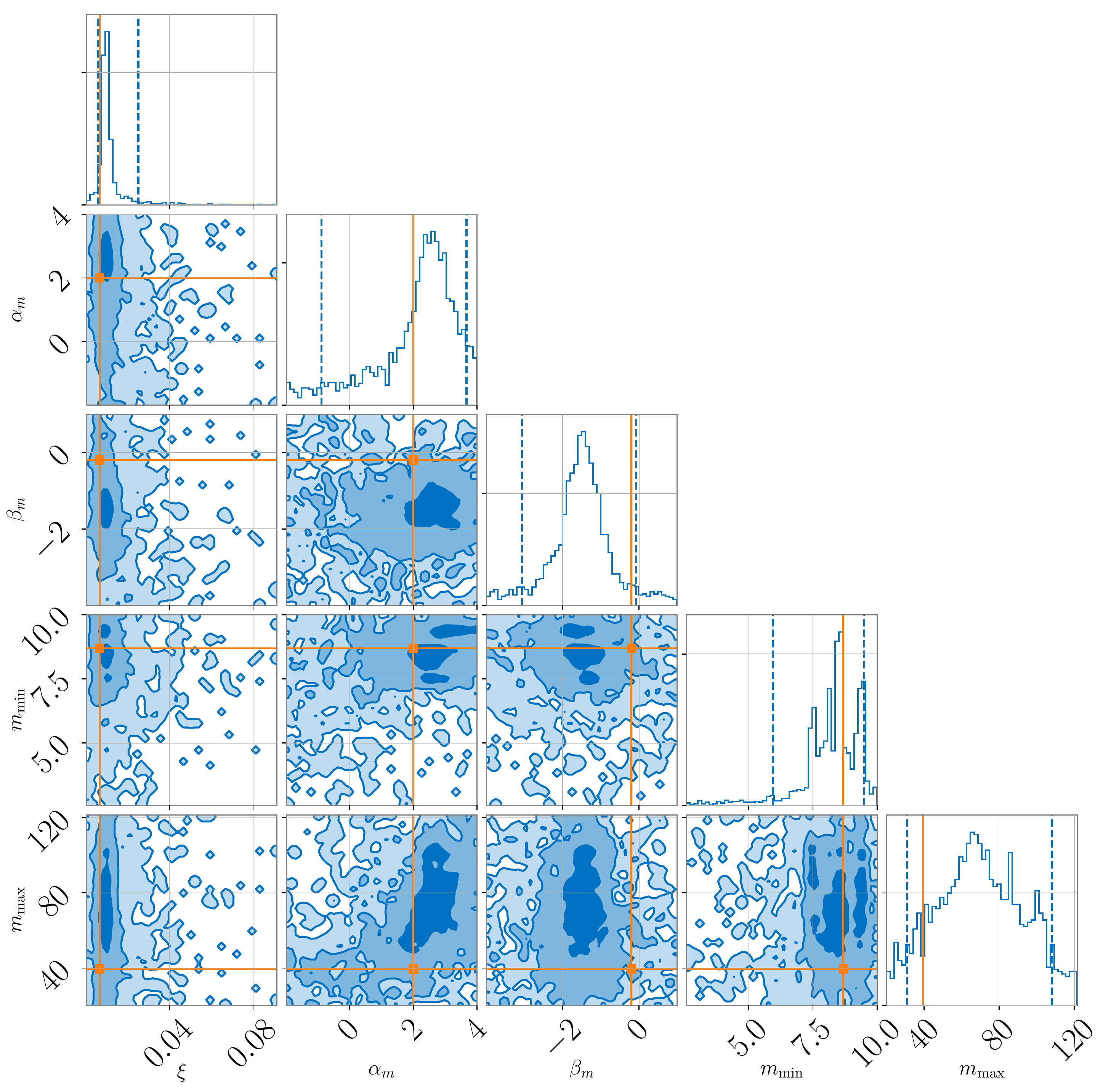}
    \caption{One- and two-dimensional (hyper-) posterior distribution. This figure showcases duty cycle $\xi$ and hyper-parameters related to the mass-spectrum peak. From left to right; the astrophysical duty cycle $\xi$; the slope of the power law component of the primary mass distribution $\alpha_m$; the slope of the mass ratio distribution $\beta_m$; the minimum black hole mass $m_{\min}$; and the maximum black hole mass in the power-law compoent $m_{\max}$. The dashed lines are the 90$\%$ credible intervals.
    }
    \label{fig:ppsn}
\end{figure*}

\begin{figure*}
    \includegraphics[width=1\textwidth]{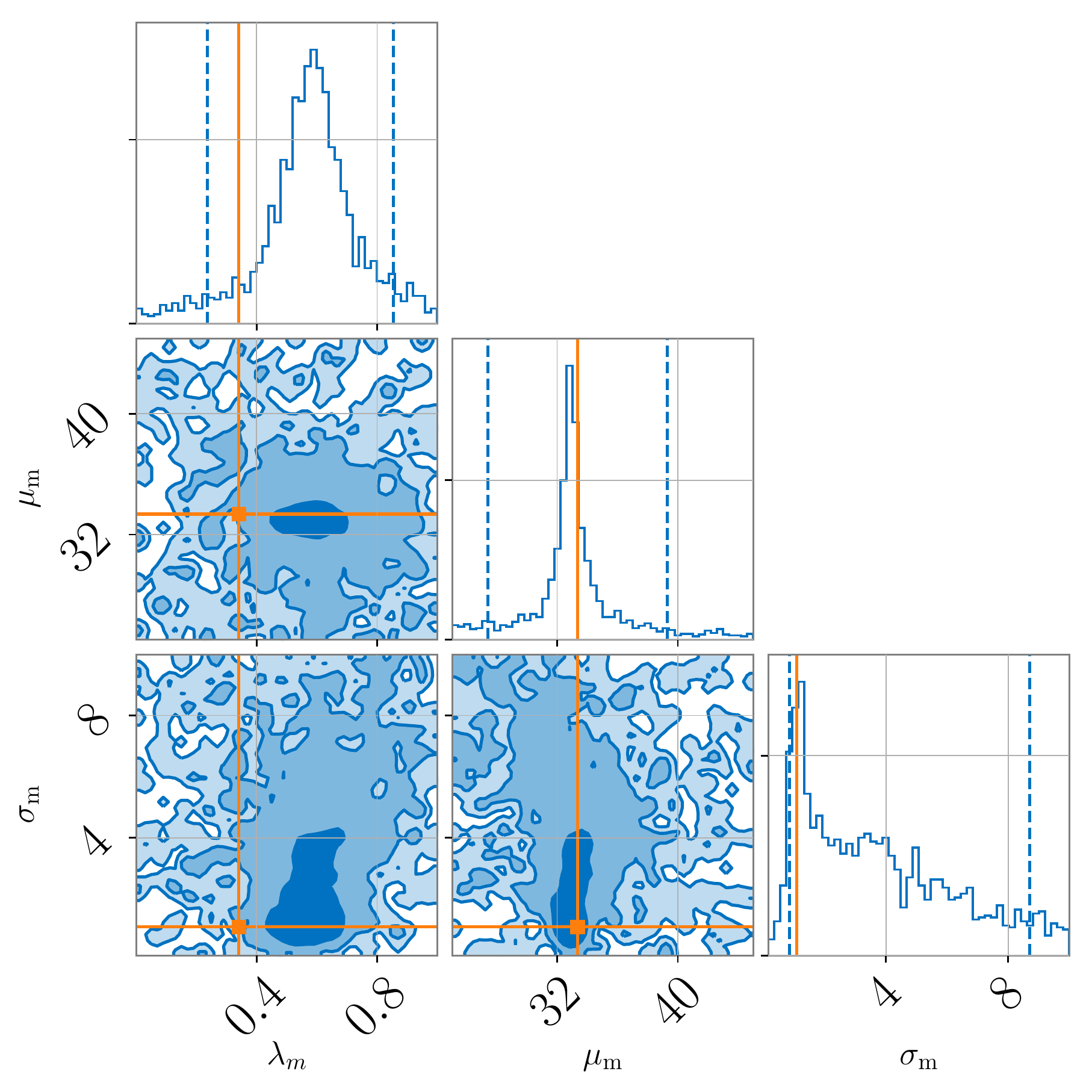}
    \caption{One- and two-dimensional (hyper-) posterior distributions. This figure showcases duty cycle  hyper-parameters related to shape of the binary black hole mass spectrum. From left to right: The fraction of black holes in the Gaussian component of the primary mass distribution $\lambda_m$; the mean of the Gaussian component of the primary mass distribution $\mu_m$; and the standard deviation of the Gaussian component of the primary mass distribution $\sigma_m$. The dashed lines are the 90$\%$ credible intervals.}
   \label{fig:other}
\end{figure*}

\begin{figure*}
    \includegraphics[width=1.0\textwidth]{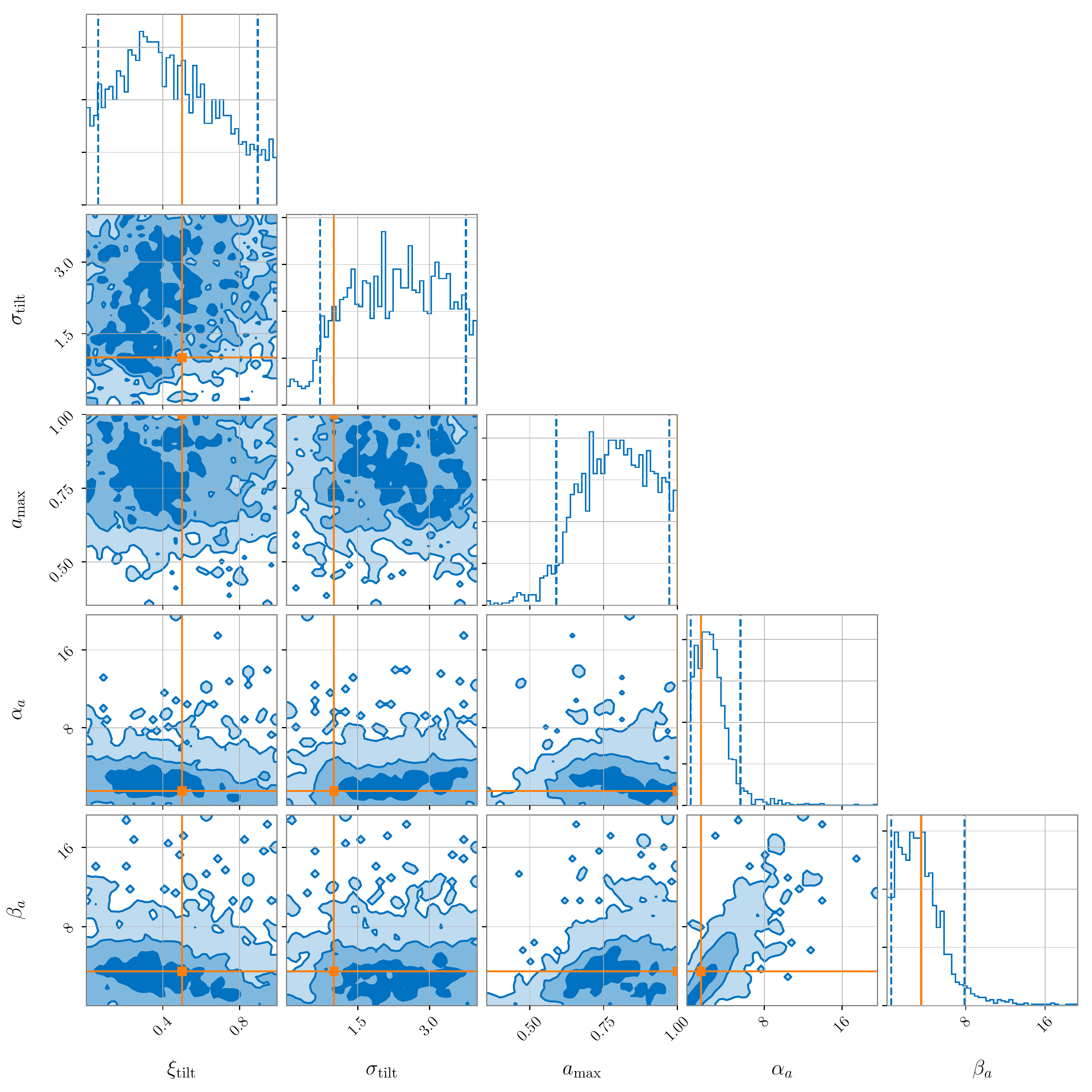}
    \caption{One- and two-dimensional (hyper-) posterior distributions. This figure shows hyper-parameters related to the distribution of black hole spins. From left the right: the  fraction of BBHs with Guassian distributed spin tilts $\xi_{\mathrm{tilt}}$; the standard deviation of the spin-tilt angle distribution $\sigma_{\mathrm{tilt}}$; the maximum spin magnitude $a_{\max}$; the spin-magnitude beta distribution slope parameter (rise) $\alpha_a$; and the spin-magnitude beta distribution slope parameter (fall) $\beta_a$.
    The dashed lines are the 90$\%$ credible intervals.
    }
    \label{fig:spins}
\end{figure*}

\section{Population model details}
\subsection{Source-frame mass}
The conditional prior for binary black hole mass is:
\begin{widetext}
\begin{equation}
\begin{split}
\label{eq:mass-model-full}
    &\pi_m(m_1 | \Lambda) =
    \left[ (1 - \lambda_{m}) A(\Lambda) \, m_{1}^{-\alpha} \, \Theta(m_{\max}-m_1) +  \lambda_{m} \, B(\Lambda) \exp\left(-\frac{(m_1-\mu_{m})^2}{2\sigma_{m}^2}\right) \right] S(m_1 | m_{\min},\delta m), \\
    &\pi_q(q | m_1, \Lambda) =
    C(m_1, \Lambda) \, q^{\beta} \, S(m_2|m_{\min},\delta m) . \\
\end{split}
\end{equation}
\end{widetext}
The first equation describes the prior probability of the primary mass $m_1$ (corresponding to the heavier of the two black holes in a binary black hole) given the hyper-parameters $\Lambda$.
The second equation describes the prior probability of the mass ratio $q=m_2/m_1$ given $m_1$ and $\Lambda$.

The fraction of black holes in the Gaussian component is $\lambda_m$.
The distribution of mass ratios follows a power-law distribution with unknown spectral index $\beta$.
Additionally, there is a smoothing parameter $\delta_{m}$ which enables the distribution to have a smooth turn-on at low masses.

The prior for primary mass $\pi(m_1|\Lambda)$ is constructed from two pieces.
The first term
\begin{align}
    (1 - \lambda_{m}) A(\Lambda) \, m_{1}^{-\alpha} \, \Theta(m_{\max}-m_1) ,
\end{align}
describes a power-law distribution with index $\alpha\in\Lambda$.
The Heaviside step-function cuts off the distribution at $m_\text{max}\in\Lambda$.
One minus the term $\lambda_m\in\Lambda$ is the fraction of events that are part of this power-law distribution.
The term $A(\Lambda)$ is a normalization constant.
This term is motivated by the fact that the stellar mass function is power-law distributed as well as evidence of a cut-off in the black hole mass spectrum~\cite{Maya,T&T17,O2R&P}.

The second term in $\pi(m_1|\Lambda)$ 
\begin{align}
    \lambda_{m} \, B(\Lambda) \exp\left(-\frac{(m_1-\mu_{m})^2}{2\sigma_{m}^2}\right) ,
\end{align}
corresponds to a Gaussian distribution with mean $\mu_m\in\Lambda$ and width $\sigma_m\in\Lambda$.
The fraction of events that are part of the Gaussian distribution is given by $\lambda_m$.
The $B(\Lambda)$ term is a normalization constant.
This term is motivated by the possibility of a bump in the black hole mass spectrum from pulsational pair instability supernovae~\cite{T&T18,O2R&P,Marchant}.

To the far right of the expression for $\pi(m_1|\Lambda)$ is a third term
\begin{align}
    \begin{split}
        S(m, m_{\min}, \delta m) &= \left( \exp f(m-m_{\min}, \delta m) + 1 \right)^{-1} \\
        f(m, \delta m) &= \frac{\delta m}{m} - \frac{\delta m}{m - \delta m}.
    \end{split}
\end{align}

The $m_\text{min}$ parameter enforces a minimum black hole mass and $\delta m$ is the mass range over which the black hole mass spectrum falls to zero.
This term is motivated by the fact that there is likely a minimum black hole mass, at least for black holes made through stellar collapse~\cite{T&T18}.

The conditional prior for mass ratio is described by a power law with index $\beta\in\Lambda$.
The smoothing function $S$ applies a low-mass cut-off in the secondary mass $m_2$, again using minimum mass $m_\text{min}$ and $\delta m$ for the mass range over which the mass spectrum falls to zero.
The variable $C(m_1,\Lambda)$ is a normalization constant.

\subsection{Lab-frame mass}
{The binary black hole lab-frame mass is a function of redshift because}
\begin{align}
    m_l = (1+z) m_s ,
\end{align}
where $m_s$ is the source-frame mass and $m_l$ is the lab-frame mass. 
When considering events at cosmological distances, the prior distributions for lab-frame masses become covariant with luminosity distance $D_L$ due to cosmological redshift.
In the source frame, the distributions of black hole mass and redshift are separable so that 
\begin{equation}
    \pi(m_{\text{s}}, z) = \pi_m(m_\text{s})\pi_z(z)
\end{equation}
Whatever form the distributions we choose for $\pi_z(z)$ and $\pi_m(m_s)$, they imply some prior for the lab-frame mass:
\begin{align}
\label{eq:mass_z_transform}
\pi(z, m_l) = & \pi\big(z, m_s(m_l)\big) 
\left|\frac{dm_s}{dm_l}\right| \nonumber\\
= & (1+z)^{-1} \pi\big(z, m_l/(1+z)\big) .
\end{align}

\subsection{Spin}
The distribution of spin magnitudes $(a_1,a_2)$ are assumed to each follow a beta distribution described by three parameters $(\alpha_{a}$, $\beta_{a}, a_\text{max})\in\Lambda$.
By treating $a_\text{max}$ as a free parameter, our model is a generalization of the prescription from~\cite{Wysocki18}.
The conditional prior for spin magnitude is
\begin{equation}
    \pi_{a}(a | \alpha_{a}, \beta_{a}, a_\text{max}) = \frac{a^{(\alpha_{a}-1)}(a_{\max}-a)^{(\beta_{a}-1)}}{a_{\max}^{(\alpha_{a}+\beta_{a}-1)}\mathrm{B}(\alpha_{a}, \beta_{a})} .
\label{eq:spin_mag}
\end{equation}
Here $\mathrm{B}(\alpha_{a},\beta_{a})$ is the Beta function.

We characterize the black hole spin orientation in terms of the cosine of the polar angle between the orbital angular momentum and the black hole spin $z_{1,2}\equiv\cos(t_{1,2})$ where $t_{1,2}$ is the polar angle .
We ignore the azimuthal angle, which has a comparatively small effect on the gravitational waveform.
We assume that the distribution of spin orientations is a mixture of an isotropic component and a preferentially aligned component modeled as a truncated half-Gaussian with unknown width $\sigma_{\text{tilt}}$ and which peaks at $t_1=t_2=1$.
\begin{equation}
\begin{split}
\label{eq:spin-alignment}
    \pi(z_1, z_2 | & {} \sigma_{\text{tilt}}, \lambda_{\text{tilt}}) =  \frac{(1-\lambda_{\text{tilt}})}{4} \\
 & {} + \frac{\lambda_{\text{tilt}}}{2\pi} \prod_{i \in \{1, 2\}} \frac{e^{-(1-z_i)^2 / (2\sigma_{\text{tilt}}^2)}}{\sigma_{\text{tilt}} \text{erf}(\sqrt{2}/\sigma_{\text{tilt}})}.
\end{split}
\end{equation}

The isotropic distribution is a model for mergers in dense stellar environments such as globular clusters, where spin orientations are expected to be isotropically oriented.
The aligned distribution models binaries formed in the field.
The fraction of binaries in the preferentially aligned component is $\xi_\chi$.
We assume that both component spins are independently drawn from the same distribution.

\bibliographystyle{mnras}
\bibliography{bib}

\begin{thebibliography}{}
\makeatletter
\relax
\def\mn@urlcharsother{\let\do\@makeother \do\$\do\&\do\#\do\^\do\_\do\%\do\~}
\def\mn@doi{\begingroup\mn@urlcharsother \@ifnextchar [ {\mn@doi@}
  {\mn@doi@[]}}
\def\mn@doi@[#1]#2{\def\@tempa{#1}\ifx\@tempa\@empty \href
  {http://dx.doi.org/#2} {doi:#2}\else \href {http://dx.doi.org/#2} {#1}\fi
  \endgroup}
\def\mn@eprint#1#2{\mn@eprint@#1:#2::\@nil}
\def\mn@eprint@arXiv#1{\href {http://arxiv.org/abs/#1} {{\tt arXiv:#1}}}
\def\mn@eprint@dblp#1{\href {http://dblp.uni-trier.de/rec/bibtex/#1.xml}
  {dblp:#1}}
\def\mn@eprint@#1:#2:#3:#4\@nil{\def\@tempa {#1}\def\@tempb {#2}\def\@tempc
  {#3}\ifx \@tempc \@empty \let \@tempc \@tempb \let \@tempb \@tempa \fi \ifx
  \@tempb \@empty \def\@tempb {arXiv}\fi \@ifundefined
  {mn@eprint@\@tempb}{\@tempb:\@tempc}{\expandafter \expandafter \csname
  mn@eprint@\@tempb\endcsname \expandafter{\@tempc}}}

\bibitem[\protect\citeauthoryear{Aasi et~al.}{Aasi et~al.}{2015}]{aligo}
Aasi J.,  et~al., 2015, Class. Quant. Grav., 32, 074001

\bibitem[\protect\citeauthoryear{Abbott et~al.}{Abbott et~al.}{2016a}]{O1_bbh}
Abbott B.~P.,  et~al., 2016a, Phys. Rev. X, 6, 041015

\bibitem[\protect\citeauthoryear{Abbott et~al.}{Abbott
  et~al.}{2016b}]{GW150914_stoch}
Abbott B.~P.,  et~al., 2016b, Phys. Rev. Lett., 116, 131102

\bibitem[\protect\citeauthoryear{Abbott et~al.}{Abbott
  et~al.}{2018a}]{GW170817_stoch}
Abbott B.~P.,  et~al., 2018a, Phys. Rev. Lett., 120, 091101

\bibitem[\protect\citeauthoryear{Abbott et~al.}{Abbott
  et~al.}{2018b}]{PhysRevLett.121.161101}
Abbott B.~P.,  et~al., 2018b, \mn@doi [Phys. Rev. Lett.]
  {10.1103/PhysRevLett.121.161101}, 121, 161101

\bibitem[\protect\citeauthoryear{Abbott et~al.}{Abbott et~al.}{2019}]{gwtc-1}
Abbott B.~P.,  et~al., 2019, Phys. Rev. X, 9, 031040

\bibitem[\protect\citeauthoryear{Abott et~al.}{Abott et~al.}{2018a}]{O2R&P}
Abott B.~P.,  et~al., 2018a

\bibitem[\protect\citeauthoryear{Abott et~al.}{Abott et~al.}{2018b}]{O2}
Abott B.~P.,  et~al., 2018b

\bibitem[\protect\citeauthoryear{Acernese et~al.,}{Acernese et~al.}{2014}]{adv}
Acernese F.,  et~al., 2014, \mn@doi [Classical and Quantum Gravity]
  {10.1088/0264-9381/32/2/024001}, 32, 024001

\bibitem[\protect\citeauthoryear{Ade et~al.,}{Ade et~al.}{2016}]{Planck_params}
Ade P. A.~R.,  et~al., 2016, \mn@doi [Astronomy & Astrophysics]
  {10.1051/0004-6361/201525830}, 594, A13

\bibitem[\protect\citeauthoryear{Akutsu et~al.}{Akutsu et~al.}{2019}]{kagra}
Akutsu T.,  et~al., 2019, \mn@doi [Nature Astronomy]
  {10.1038/s41550-018-0658-y}, 3, 35–40

\bibitem[\protect\citeauthoryear{Ashton et~al.}{Ashton et~al.}{2018}]{bilby}
Ashton G.,  et~al., 2018

\bibitem[\protect\citeauthoryear{Farr, Stevenson, Miller, Mandel, Farr  \&
  Vecchio}{Farr et~al.}{2017}]{FarrNature}
Farr W.~M.,  Stevenson S.,  Miller M.~C.,  Mandel I.,  Farr B.,   Vecchio A.,
  2017, Nature, 548, 426

\bibitem[\protect\citeauthoryear{Fishbach \& Holz}{Fishbach \&
  Holz}{2017}]{mass_uc}
Fishbach M.,  Holz D.~E.,  2017, Astrophys. J. Lett., 851, L25

\bibitem[\protect\citeauthoryear{Fishbach, Holz  \& Farr}{Fishbach
  et~al.}{2017}]{Maya}
Fishbach M.,  Holz D.,   Farr B.,  2017, Astrophys. J. Lett., 840, L24

\bibitem[\protect\citeauthoryear{Fishbach, Holz  \& Farr}{Fishbach
  et~al.}{2018}]{Maya2}
Fishbach M.,  Holz D.~E.,   Farr W.~M.,  2018, Astrophys. J. Lett., 863, L41

\bibitem[\protect\citeauthoryear{Gaebel, Veitch, Dent  \& Farr}{Gaebel
  et~al.}{2018}]{Gaebel}
Gaebel S.~M.,  Veitch J.,  Dent T.,   Farr W.~M.,  2018

\bibitem[\protect\citeauthoryear{Gerosa \& Berti}{Gerosa \&
  Berti}{2017}]{GerosaBerti}
Gerosa D.,  Berti E.,  2017, Phys. Rev. D, 95, 124046

\bibitem[\protect\citeauthoryear{Hannam, Schmidt, Boh{\'{e}}, Haegel, Husa,
  Ohme, Pratten  \& P{\"{u}}rrer}{Hannam et~al.}{2014}]{IMRPhenomP}
Hannam M.,  Schmidt P.,  Boh{\'{e}} A.,  Haegel L.,  Husa S.,  Ohme F.,
  Pratten G.,   P{\"{u}}rrer M.,  2014, Phys. Rev. Lett., 113, 151101

\bibitem[\protect\citeauthoryear{Hernandez-Vivanco, Smith, Thrane  \&
  Lasky}{Hernandez-Vivanco et~al.}{2019}]{tbs-bns}
Hernandez-Vivanco F.,  Smith R. J.~E.,  Thrane E.,   Lasky P.~D.,  2019, Phys.
  Rev. D, 100, 043023

\bibitem[\protect\citeauthoryear{Lower, Thrane, Lasky  \& Smith}{Lower
  et~al.}{2018}]{eccentricity}
Lower M.~E.,  Thrane E.,  Lasky P.~D.,   Smith R.,  2018, Phys. Rev. D, 98,
  083028

\bibitem[\protect\citeauthoryear{Madau \& Dickinson}{Madau \&
  Dickinson}{2014}]{madauSFR}
Madau P.,  Dickinson M.,  2014, \mn@doi [Annual Review of Astronomy and
  Astrophysics] {10.1146/annurev-astro-081811-125615}, 52, 415

\bibitem[\protect\citeauthoryear{Maggiore}{Maggiore}{2000}]{Maggiore}
Maggiore M.,  2000, Phys. Rep., 331, 283

\bibitem[\protect\citeauthoryear{Mandel, Farr  \& Gair}{Mandel
  et~al.}{2018}]{MandelFarrGair}
Mandel I.,  Farr W.~M.,   Gair J.~R.,  2018

\bibitem[\protect\citeauthoryear{Marchant, Renzo, Farmer, Pappas, Taam, de Mink
   \& Kalogera}{Marchant et~al.}{}]{Marchant}
Marchant P.,  Renzo M.,  Farmer R.,  Pappas K. M.~W.,  Taam R.~E.,  de Mink S.,
    Kalogera V.,

\bibitem[\protect\citeauthoryear{Ng, Vitale, Zimmerman, Chatziioannou, Gerosa
  \& Haster}{Ng et~al.}{2018}]{Ng}
Ng K. K.~Y.,  Vitale S.,  Zimmerman A.,  Chatziioannou K.,  Gerosa D.,   Haster
  C.-J.,  2018, Phys. Rev. D, 98, 083007

\bibitem[\protect\citeauthoryear{Raidal, Vaskonen  \& VeermÃ€e}{Raidal
  et~al.}{2017}]{1475-7516-2017-09-037}
Raidal M.,  Vaskonen V.,   VeermÃ€e H.,  2017, Journal of Cosmology and
  Astroparticle Physics, 2017, 037

\bibitem[\protect\citeauthoryear{Smith \& Thrane}{Smith \& Thrane}{2018}]{tbs}
Smith R.,  Thrane E.,  2018, \mn@doi [Phys. Rev. X]
  {10.1103/PhysRevX.8.021019}, 8, 021019

\bibitem[\protect\citeauthoryear{Smith, Field, Blackburn, Haster, P{\"{u}}rrer,
  Raymond  \& Schmidt}{Smith et~al.}{2016}]{PhysRevD.94.044031}
Smith R.,  Field S.~E.,  Blackburn K.,  Haster C.-J.,  P{\"{u}}rrer M.,
  Raymond V.,   Schmidt P.,  2016, Phys. Rev. D, 94, 44031

\bibitem[\protect\citeauthoryear{Speagle}{Speagle}{2020}]{dynesty}
Speagle J.~S.,  2020, \mn@doi [Monthly Notices of the Royal Astronomical
  Society] {10.1093/mnras/staa278}, 493, 3132–3158

\bibitem[\protect\citeauthoryear{Stevenson, Ohme  \& Fairhurst}{Stevenson
  et~al.}{2015}]{Stevenson}
Stevenson S.,  Ohme F.,   Fairhurst S.,  2015, Astrophys. J., 810, 58

\bibitem[\protect\citeauthoryear{Stevenson, Berry  \& Mandel}{Stevenson
  et~al.}{2017}]{Stevenson2}
Stevenson S.,  Berry C.,   Mandel I.,  2017, MNRAS, 471, 2801

\bibitem[\protect\citeauthoryear{Talbot \& Thrane}{Talbot \&
  Thrane}{2017a}]{spin}
Talbot C.,  Thrane E.,  2017a, Phys. Rev. D, 96, 023012

\bibitem[\protect\citeauthoryear{Talbot \& Thrane}{Talbot \&
  Thrane}{2017b}]{T&T17}
Talbot C.,  Thrane E.,  2017b, \mn@doi [Phys. Rev. D]
  {10.1103/PhysRevD.96.023012}, 96, 023012

\bibitem[\protect\citeauthoryear{Talbot \& Thrane}{Talbot \&
  Thrane}{2018}]{T&T18}
Talbot C.,  Thrane E.,  2018, The Astrophysical Journal, 856, 173

\bibitem[\protect\citeauthoryear{Thrane \& Talbot}{Thrane \&
  Talbot}{2018}]{ThraneTalbot}
Thrane E.,  Talbot C.,  2018

\bibitem[\protect\citeauthoryear{Tiwari, Fairhurst  \& Hannam}{Tiwari
  et~al.}{2018}]{Tiwari}
Tiwari V.,  Fairhurst S.,   Hannam M.,  2018, Astrophys. J., 868, 140

\bibitem[\protect\citeauthoryear{Vitale, Lynch, Sturani  \& Graff}{Vitale
  et~al.}{2017}]{salvo}
Vitale S.,  Lynch R.,  Sturani R.,   Graff P.,  2017, Class. Quant. Grav., 34,
  03LT01

\bibitem[\protect\citeauthoryear{Wysocki, Lange  \& O.~'shaughnessy}{Wysocki
  et~al.}{2018}]{Wysocki18}
Wysocki D.,  Lange J.,   O.~'shaughnessy R.,  2018

\bibitem[\protect\citeauthoryear{You, Zhu, Ashton, Thrane  \& Zhu}{You
  et~al.}{2020}]{standard-siren}
You Z.-Q.,  Zhu X.-J.,  Ashton G.,  Thrane E.,   Zhu Z.-H.,  2020

\makeatother
\end{thebibliography}

\end{document}